\documentclass{optica-article}

\journal{oe}

\articletype{Research Article}

\usepackage{lineno}

\usepackage[separate-uncertainty=true,multi-part-units=single]{siunitx}
\usepackage{float}

\DeclareSIUnit{\pix}{\text{pixel}}

\usepackage[acronym]{glossaries}
\newacronym{scidar}{SCIDAR}{SCIntillation Detection And Ranging}
\newacronym{slodar}{SLODAR}{SLOpe Detection And Ranging}
\newacronym{shimm}{24hSHIMM}{24-hour Shack-Hartmann Image Motion Monitor}
\newacronym{dimm}{DIMM}{Differential Image Motion Monitor}
\newacronym{ao}{AO}{Adaptive Optics}
\newacronym{wfs}{WFS}{Shack-Hartmann wavefront sensor}
\newacronym{int}{INT}{Isaac Newton Telescope}
\newacronym{wht}{WHT}{William Herschel Telescope}
\newacronym{ingaas}{InGaAs}{Indium Gallium Arsenide}
\newacronym{cad}{CAD}{Computer-Aided Design}
\newacronym{gsm}{GSM}{Generalised Seeing Monitor}
\newacronym{nir}{NIR}{near-infrared}
\newacronym{swir}{SWIR}{short-wave infrared}
\newacronym{fsoc}{FSOC}{Free Space Optical Communications}
\newacronym{mass}{MASS}{Multi-Aperture Scintillation Sensor}
\newacronym{cmos}{CMOS}{Complementary Metal-Oxide Semiconductor}
\newacronym{ssa}{SSA}{Space Situational Awareness}
\newacronym{sst}{SST}{Space Surveillance and Tracking}
\newacronym{qkd}{QKD}{Quantum Key Distribution}
\newacronym{nnls}{NNLS}{Non-Negative Least Squares}
\newacronym{shabar}{SHABAR}{SHAdow BAnd Ranging}
\newacronym{pml}{PML}{Profiler of the Moon Limb}
\newacronym{sdimm}{S-DIMM}{Solar-DIMM}

\newcommand{\comment}[1]{}
\newcommand{\cn}{\ensuremath{C_n^2(h)\:\!\mathrm{d}h}} 
\newcommand{\review}[1]{{#1}}

\begin{document}

\title{Demonstrating 24-hour continuous vertical monitoring of atmospheric optical turbulence}

\author{Ryan Griffiths\authormark{1,*}, James Osborn\authormark{1}, Ollie Farley\authormark{1}, Tim Butterley\authormark{1}, Matthew J. Townson\authormark{1} and Richard Wilson\authormark{1}}

\address{\authormark{1}Department of Physics, Centre for Advanced Instrumentation, Durham University, UK, DH1 3LE}

\email{\authormark{*}ryan.griffiths@durham.ac.uk}

\begin{abstract*}
We report what is believed to be the first example of fully continuous, 24-hour vertical monitoring of atmospheric optical turbulence. This is achieved using a novel instrument, the \gls{shimm}. Optical turbulence is a fundamental limitation for applications such as free-space optical communications, where it limits the achievable bandwidth, and ground-based optical astronomy, restricting the observational precision. Knowledge of the turbulence enables us to select the best sites, design optical instrumentation and optimise the operation of ground-based optical systems.

\review{The \gls{shimm} estimates the vertical optical turbulence coherence length, time, angle and Rytov variance from the measurement of a four-layer vertical turbulence profile and a wind speed profile retrieved from meteorological forecasts. To illustrate our advance we show the values of these parameters recorded during a 35-hour, continuous demonstration of the instrument.}
Due to its portability and ability to work in stronger turbulence, the 24hSHIMM can also operate in urban locations, \review{providing the field with a truly continuous, versatile turbulence monitor for all but the most demanding of applications.}
\end{abstract*}
\comment{\review{We show that this instrument enables truly continuous day and night  monitoring of atmospheric optical turbulence parameters, providing the field with a versatile monitor for all but the most demanding of applications.}}

\glsresetall

\section{Introduction}

The turbulence in the Earth's atmosphere affects the propagation of light by causing images of point sources to appear to rapidly change in brightness (scintillation) and break up into speckles. This optical turbulence is caused by the mixing of air of different temperatures, and hence density, developing spatial and temporal variations in refractive index. Atmospheric optical turbulence is a major limitation to several mature and emerging applications, such as; \gls{fsoc} \cite{Chan2006,Matacalvo19}, either horizontally or to and from space and including long-range \gls{qkd}\cite{Erven2008,Polnik2020}; imaging of space objects for \gls{sst} \cite{Copeland2016}; space debris deorbiting via acceleration due to laser ablation or photon pressure \cite{Bennet2012}; and both daytime and night-time ground-based astronomy (for example,  \cite{Roddier1981}).

For each of these applications knowing the strength and variability of the optical turbulence enables us to select the best possible sites, model how each system can perform, optimise the optical design\comment{\cite{Devaney1998,Schoeck09}}, and develop novel turbulence mitigation schemes, for example using \gls{ao} technology.

Continuous 24-hour turbulence monitoring is required to support \gls{fsoc} and \gls{sst}, where facilities will need to operate continuously, night and day. \gls{fsoc} will also need to operate in urban and suburban locations, environments that are likely to have stronger turbulence than the pristine mountain top sites of most astronomical observatories where turbulence monitoring instrumentation tends to be located.

\comment{
In addition, accelerating development in flagship projects such as the European Space Agency's European Data Relay Service \cite{Calzolaio2020}, growing interest in the development of \gls{ao} systems for optical ground stations \cite{Sodnik2012,Fischer2021,Osborn2021} and high performance for applications such as \gls{qkd} \cite{Gruneisen2021}, there is more demand than ever for 24-hour turbulence monitoring systems. 
}

Even for astronomy the ability to measure the turbulence continuously enables \comment{an estimate of the conditions before the start of observations, but also enable }data assimilation into turbulence forecasting tools. This continuous data assimilation has been shown to significantly improve the performance of turbulence forecasting tools enabling a much more efficient smart scheduling of observations \cite{Masciadri2019}, ensuring that the most sensitive observations are carried out at the optimum time, maximising the probability of success.

Here, we demonstrate for the first time that a system that exploits a \gls{wfs} operating in the \gls{swir} and utilises techniques for wavefront-sensing in high noise can provide continuous 24-hour vertical turbulence monitoring even in strong turbulence conditions. The system is built around a commercial off-the-shelf 11" telescope and the low weight and size make it ideal for portable campaigns anywhere in the world. The instrument, which we call the \gls{shimm}, monitors a single bright star and can extract the atmospheric turbulence coherence length, $r_0$, coherence angle, $\theta_0$, coherence time, $\tau_0$, the Rytov variance, $\sigma_R^2$, and a four-layer vertical profile. This system is a development of the SHIMM\cite{Perera2017}. However, until now, the SHIMM has been limited to night-time operations only. The work presented here describes a significant upgrade to existing systems in both the hardware and analysis and represents a necessary advance in the field of turbulence monitoring.
Current techniques are limited to either day or night-time, leading to differential instrument bias between the two windows and a significant gap in measurement during twilight hours. They are also generally restricted to weak turbulence, limiting their use for monitoring conditions in urban environments, for example near data and population centres, a requirement for free-space optical communications networks.

\section{The 24hSHIMM}

\begin{figure}
     \includegraphics[height=5cm]{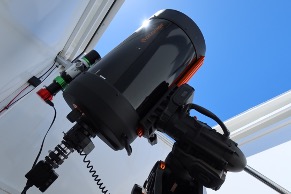}
    \centering
    \includegraphics[height=5.1cm]{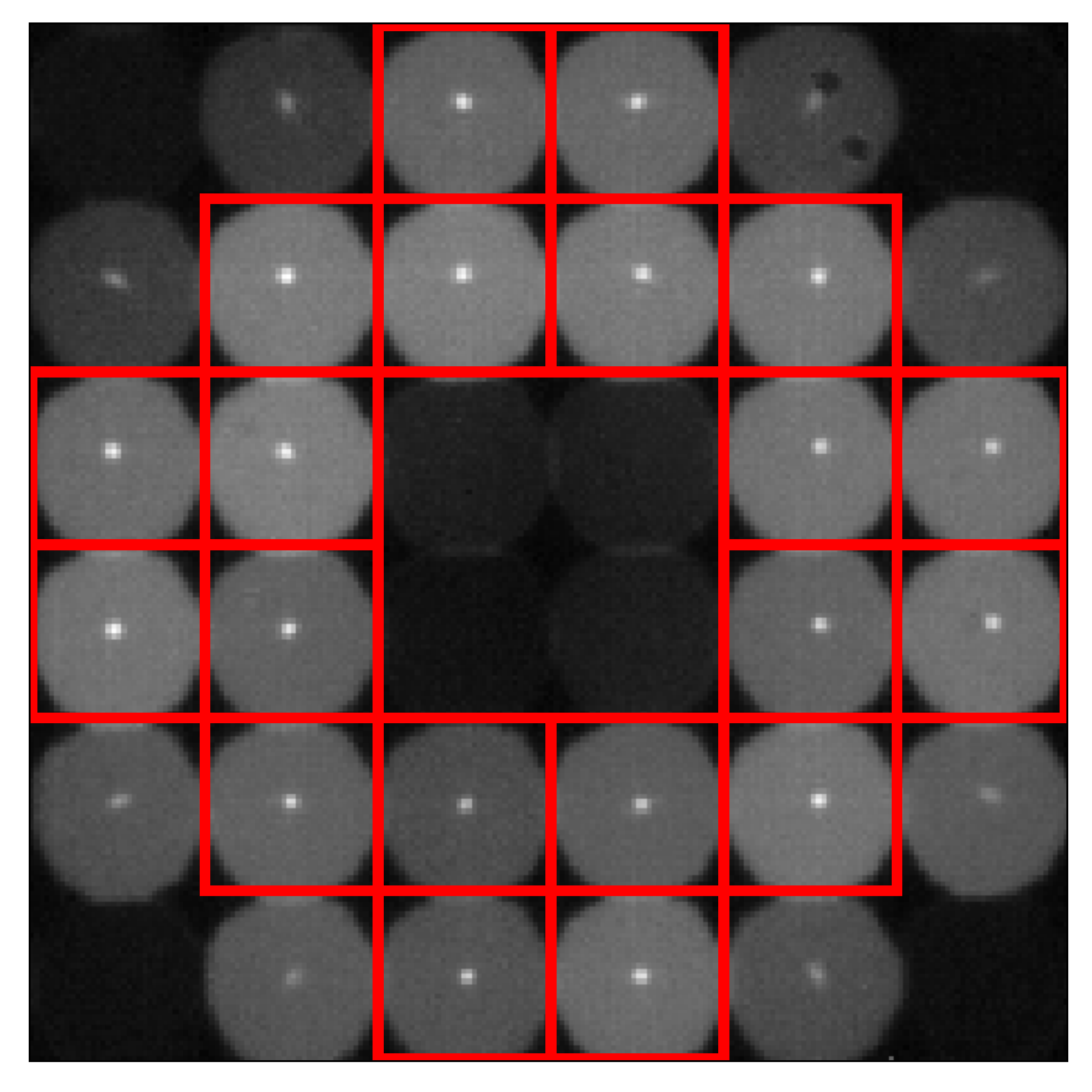}
    \caption{Left, a photo of the \gls{shimm} operating during daytime on La Palma, Spain. Right, a typical \gls{wfs} frame from the \gls{shimm} taken at approximately 2pm local time in La Palma. The red squares indicate the sub-aperture focal spots used for data analysis.}
    \label{fig:wfs}
\end{figure}

\begin{figure}
    \centering
    \includegraphics[width=0.65\textwidth,trim={5cm 1cm 5cm 1cm},clip]{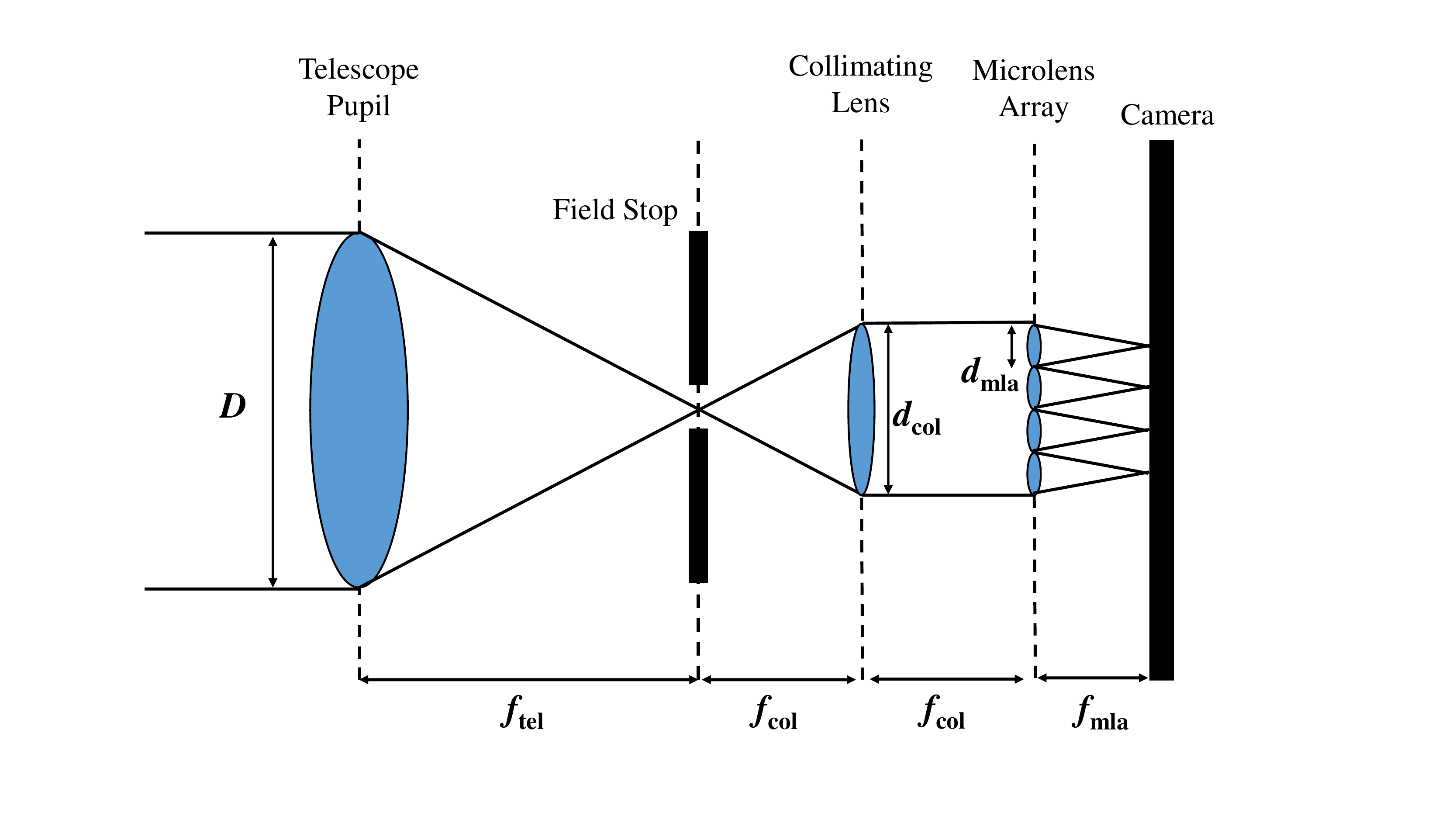} 
    \caption{Simple \gls{shimm} optical design. $f_\mathrm{tel},f_\mathrm{col},f_\mathrm{mla}$ are the focal lengths of the telescope, collimating lens and microlens array respectively and the $d_\mathrm{col},d_\mathrm{mla}$ and $D$ are the diameters of the collimated beam, the microlens array lenslets and the telescope pupil respectively.}
    \label{fig:design}
\end{figure}

In selecting equipment for the \gls{shimm}, an effort was made to maintain portability while maximising the contrast of daytime \gls{wfs} images with the vast majority of equipment obtained off-the-shelf. Fig.~\ref{fig:wfs} shows the assembled instrument. For the telescope a Celestron C11 279.4mm SCT \comment{telescope} was chosen to maximise mirror collecting area while retaining portability and ease of setup. This was paired with a low-cost but programmable Celestron CGX mount. The daytime brightness of the sky background decreases steeply with increasing wavelength as a result of Rayleigh scattering, therefore optical components were optimised for \gls{swir} observations. For autoguiding, a SkyWatcher Evoguide 50ED guidescope was paired with a ZWO ASI462MC CMOS camera (60\% QE at \SI{850}{\nm}) and an \SI{850}{\nm} longpass filter to block visible wavelengths. The \gls{wfs} detector was a First Light Imaging CRED-3 \gls{ingaas} camera \cite{Gach2020} with a bandpass of 900-\SI{1700}{\nm}. The optical design of the \gls{shimm} is shown in Fig.~\ref{fig:design}. We used a NIR-II coated achromatic doublet for the collimating lens, \SI{500}{\micro\meter}-pitch square lenslet array and a \SI{0.8}{\mm} minimum diameter (corresponding to a minimum field-of-view of \SI{59}{\arcsecond}) iris to function as a field stop minimising background light. The collimating lens and microlens array focal lengths were chosen to produce six \SI{4.7}{\cm}-wide sub-apertures across the pupil. This microlens array sub-aperture size coupled with our choice of camera ensured that \gls{wfs} focal spots were at least Nyquist-sampled and motions due to wavefront distortion could be measured accurately. This was verified in simulation as detailed in section \review{4.1}. \comment{A photo of the instrument is shown in Fig.~\ref{fig:wfs} along with an example image of the lenslet spots on the detector.}

\comment{Two designs were produced for the \gls{shimm}. One utilising a low-cost ZWO ASI174MM CMOS camera with a \SI{600}{\nm} filter yielding an effective bandpass of 600-\SI{1100}{\nm} and the other a First Light Imaging CRED-3 \gls{ingaas} camera \cite{Gach2020} with a bandpass of 900-\SI{1700}{\nm}. Common to both designs were an appropriately coated achromatic doublet for the collimating lens, \SI{500}{\micro\meter}-pitch square lenslet array and an iris to function as a field stop, its \SI{0.8}{\mm} minimum diameter corresponding to a \SI{59}{\arcsecond} field-of-view in each sub-aperture. The optical design of the \gls{shimm} is shown in Fig.~\ref{fig:design}. Firstly, the field stop was placed in the focal plane of the telescope to restrict the field of view, the telescope beam was then collimated by an achromatic lens, chosen to reduce chromatic aberration from these wide bandpasses. For the CMOS design a \SI{600}{\nm} filter was placed in the collimated beam to reduce the sky brightness. Following this a microlens array was placed in the pupil plane to focus light onto the detector. Photos of the instrument are show in Fig.~\ref{fig:photos}.}

\comment{
\begin{figure}
    \centering
    \includegraphics[height=5cm]{figures/Picture 1.jpg}
    \caption{Photos of the \gls{shimm} operating during daytime on La Palma, Spain. XXX Add photo from London XXX}
    \label{fig:photos}
\end{figure}
}

The instrument was aligned with the aid of a 3d Autodesk Inventor model of the optomechanics integrating a model of the optics produced in Zemax. \comment{Vernier calipers with a measurement error of $\pm\SI{0.01}{\mm}$ were then used to position the components along a Thorlabs 30mm cage system.} 
After achieving focus, alignment errors were accounted for by performing an image scale calibration on a double star with a known angular separation. This gave \comment{To account for this, an image scale calibration on a double star was performed at night giving} an average image scale of \comment{$\rho = \SI{0.905\pm0.009}{\arcsecond\per\pix}$ compared to the predicted \SI{0.848}{\arcsecond\per\pix} for the CMOS design used in Durham, and}$\rho = \SI{2.27\pm0.01}{\arcsecond\per\pix}$ compared to the predicted  \SI{2.17}{\arcsecond\per\pix}\comment{ for the \gls{ingaas} design used in La Palma}. The effective wavelength of the system, a key variable for the calculation of atmospheric parameters, was found to be \review{\SI{1280}{\nm}. The dependence of the effective wavelength on the stellar spectrum was ignored. The value of the effective wavelength was} estimated by constructing a theoretical transmission curve based on manufacturer data for the optics and camera and a study into the Celestron corrector plate coating \cite{Baril2010} which we use as an estimate of the transmission of the three surfaces in the telescope. \review{The last assumption is likely to result in a systematic error in the effective wavelength calculation, however this was the best possible estimate in the absence of telescope transmission data at \gls{swir} wavelengths.}

The experimental setup consisted of a Linux workstation running the cameras, mount and data acquisition software and could be remotely configured and controlled. The process by which data were collected was as follows. Initially stars with a zenith angle greater than \ang{40} were filtered out and the remaining stars were ordered by J-band magnitude. The best target was then judged primarily on brightness, but also on separation from the sun and total observation time. The \gls{shimm} was then made to track the best available star and the \gls{wfs} camera set to run at a short exposure time of 2ms to ensure that wavefront measurements were not smeared over the course of the exposure \cite{Wilson2009}. \review{For a typical wind speed of \SI{20}{\meter\per\second} this will produce a blur of 4cm, which is comparable to the sub-aperture size, and bias caused by the finite exposure time is ignored.} 3600 images were recorded at approximately 120 Hz constituting a single data point or measurement. After each measurement, if necessary, the star could be guided back into the centre of the field of view using the autoguiding equipment. This process was repeated until either the target exceeded the zenith angle threshold or the mount would be forced perform a meridian flip and a new target had to be found. With a robust algorithm for determining optimal targets this process may be fully automated to gather large quantities of continuous data.

\review{To investigate target availability for the \gls{shimm}, the minimum required signal-to-noise ratio was estimated by running simulations for a variety of shot signals using detector noise parameters quoted in section 4.1. This investigation returned a shot signal of approximately 5000$e^-$ corresponding to a maximum J band magnitude of 1.5 for the target star. Using data from a star catalogue \cite{Ducati2002} and this magnitude limit, the number of valid target stars at each hour was investigated for the date and location of the 35-hour experiment in La Palma. It was found that at any given  time there would be a minimum of 20 targets available. However, due to limitations in the mount pointing accuracy, the guidescope, which was significantly more limited by sky background noise, was relied upon to acquire targets. The magnitude limit used in the experiment was zero during the day - still providing at least one usable target at all times.} 

\section{Methods}
\comment{why these parameters}
\comment{is a low resolution profile method good enough}

\subsection{Shack-Hartmann daytime image analysis}
The centroids of pixel values of the focal spots in the Shack-Hartmann lenslet array are calculated via a version of the moving window centre-of-gravity algorithm \cite{Thomas2006} using a top-hat window in conjunction with a brightest pixel centre-of-gravity calculation \cite{Basden2012} to find the centroid position within the moving window. The centroids not only give the wavefront slope across each Shack-Hartmann sub-aperture, but also enable accurate measurement of the intensity in the spots, which is here achieved by summing the pixel values in a small circle surrounding the centroid.

Without a careful background subtraction, the centroid and intensity of the Shack-Hartmann spot cannot be calculated accurately and the rapidly changing sky background at sunset and sunrise presents a significant challenge as any on-sky background images can quickly become obsolete. Background subtraction for the \gls{shimm} is therefore implemented in the real-time data analysis stage. The mean background level is measured in a thin annulus around each sub-aperture spot, which is then subtracted from each image. \review{When calculating the spot centroids, a 2x2 median filter is applied to the images to reduce centroid bias from the residual noise and and hot pixels present. This filter is not applied to images when measuring the spot intensities.} 

\review{Although the use of an \gls{ingaas} camera minimises sky background noise during the daytime, these sensors have high levels of fixed pattern noise and dark noise, especially when uncooled. Indeed during experiments with the \gls{shimm} it was observed that the combined quadrature sum of detector readout noise variance and dark noise variance was $66.5 e^- \mathrm{pix}^{-1}$. This was greater than the daytime sky background noise. The CRED-3 camera was chosen due to its high linearity and suitability for wavefront sensing applications \cite{Gach2020}. The camera is uncooled - reducing weight and expense, however by running at the very short exposure times required for wavefront sensing, the increased contribution of dark current to the detector noise as a result of higher sensor temperature is minimised. Hot pixels are calibrated through a factory-set mask provided with the camera. Furthermore to address the high levels of fixed pattern noise and dark noise, master-dark images are taken at regular intervals throughout experiments as the sensor temperature changes. Additionally the \gls{shimm} utilises a small 224x224 pixel region of the sensor minimising the small effects of sensor non-linearity on centroid measurements.}

\comment{Firstly, a mean image of each sub-aperture on the wavefront sensor is produced by calculating the average of all the frames in a measurement sequence. From this, a mean centroid position is calculated and a circular mask is applied on this position to block out the spot in the mean image. The mean background level is measured from a thin annulus around this circular mask and the pixels in the circular mask are then set to this value. This produces a "background image" that may be subtracted from every other frame in the sequence. Negative pixels are then set to zero and a 2x2 median filter is applied to the images to reduce centroid bias from the residual noise and and hot pixels present.}

\comment{\subsection{Optical Turbulence Strength}
The coherence length (or Fried parameter), $r_0$, is often used to quantify the integrated strength of the turbulence. This is a useful parameter as it is defined as the diameter of an aperture in which the phase variance is approximately one. Stronger turbulence will therefore correspond to a smaller $r_0$.}

\comment{
\begin{figure}[h!]
    \centering
    \includegraphics[width=0.9\textwidth]{figures/subap_bgs_img.pdf}
    \caption{}
    \label{fig:background}
\end{figure}
}

\subsection{Four-layer vertical profile}
The vertical profile of the optical turbulence strength, \cn, is estimated from the auto-covariance of the intensity in the Shack-Hartmann sub-apertures. This is similar to the SCO-SLIDAR (Single Coupled SLODAR SCIDAR) method \cite{Vedrenne2007,Robert2006} but only utilising the intensity covariance information, removing issues associated with telescope vibrations in the phase covariance. \review{This is also the approach taken by the Shack-Hartmann MASS instrument \cite{Ogane2021}}. Adopting here similar notation to \cite{Vedrenne2010}, the vector of measured intensity covariances is denoted $\mathbf{C}$, the unbiased covariances $\mathbf{C'}$, the matrix of weighting functions $\mathbf{M}$ and the unknown \cn\  vector $\mathbf{S}$. \review{The weighting functions $\mathbf{M}$ are computed using standard weak-scintillation theory for monochromatic light at the effective wavelength of the instrument, not accounting for the effects of finite spectral bandwidth and partial saturation}. Additionally a vector, $\mathbf{E}$, containing the bias to the intensity covariances as a result of photometric noise is introduced and the inversion problem is given by

\begin{equation}\label{eq:inverse}
 \mathbf{C} - \mathbf{E} = \mathbf{C'} = \mathbf{M}\, \mathbf{S}.
\end{equation}

It is assumed that the photometric noise is independent in each sub-aperture, and therefore that intensity covariances between different sub-apertures are unbiased and the corresponding components of $\mathbf{E}$ are equal to zero. For intensity covariances between a sub-aperture and itself, denoted components $k$ in $\mathbf{C}$, the bias as a result of photometric noise, and therefore the corresponding components $k$ in $\mathbf{E}$, are

\begin{equation}\label{eq:bias}
    \mathbf{E}_{\mathrm{k}} = \left[S + n_{\mathrm{pix}} \left( B + D + \sigma_{\mathrm{Rd}}^2 \right) \right] / \left\langle S \right\rangle ^ 2.
\end{equation}

For a single exposure, $S$ is the sub-aperture shot signal from the target star, $n_{\mathrm{pix}}$ is the number of pixels used to measure the spot intensity \review{and its typical value in this work is 29}, $B$ and $D$ are the sky background and dark current counts per pixel, $\sigma_{\mathrm{Rd}}$ is the RMS readout noise per pixel and the angular brackets denote an average over all frames in the measurement sequence. Eq.~(\ref{eq:inverse}) is then solved for $\mathbf{S}$ using a non-negative least squares algorithm. The uncertainties in the values of the reconstructed vertical turbulence profile, \cn, are estimated through the bootstrap method \cite{Efron1979BootstrapJackknife}. This involves randomly selecting (with replacement) frames from a measurement, performing the profile reconstruction from this sample and repeating this a large number of times. The standard deviation of the resulting sampled profiles is therefore the bootstrap estimation of the standard error in the reconstructed profile.

However, the key limitation of this approach is that the intensity is insensitive to the ground layer turbulence. Therefore, the ground layer strength is found by subtracting the sum of this profile from the integrated turbulence strength obtained from \gls{wfs} slopes. \review{This is generally a similar approach to \cite{Potanin2022} and the Multi-Aperture Scintillation Sensor - Differential Image Motion Monitor \cite{Kornilov2007}, except in this analysis, as explained in section 3.3, the integrated turbulence strength is found by comparing the auto-covariance of \gls{wfs} angle of arrival slope measurements with a theoretical response.} The ground layer is placed at \SI{0}{\meter} and the vertical heights of the layers reconstructed by solving Eq.~(\ref{eq:inverse}) are chosen to be 4, 12, and \SI{20}{\kilo\meter}.

\subsection{Atmospheric turbulence parameters}

The \gls{shimm} takes advantage of the \gls{slodar} angle of arrival auto-covariance analysis \cite{Wilson2002} to measure the turbulence coherence length (or Fried parameter), $r_0$, \review{which is related to the integrated turbulence strength}. This approach involves calculating the theoretical centroid auto-covariance response of the \gls{wfs} for imaging through Kolmogorov turbulence \review{of a given strength}  and fitting it to measured auto-covariances from the \gls{shimm} \cite{Perera2016,Butterley2006}. \review{The estimate of the coherence length is then further refined by scintillation correction using the four-layer vertical profile as described in section 3.4.}

The remaining atmospheric turbulence parameters are calculated from the four-layer vertical \cn\  profile. \review{Equations used to calculate the coherence angle, Rytov variance and coherence time are found in \cite{Roddier1981}. The coherence angle, $\theta_0$, represents the largest angle on-sky over which \gls{ao} corrections may be considered valid. The Rytov variance, $\sigma_R^2$, defines the magnitude of the received intensity variance after passing through the atmosphere. This is equivalent to the scintillation index for a point in weak turbulence ($\sigma_R^2<0.3$). These two parameters may be calculated directly from the measured \cn\ profile.} To estimate the optical turbulence coherence time, the vertical profile of the wind speed is also required as faster turbulence will lead to shorter coherence time. The vertical profile of the wind velocity, $V(h)$ is extracted from the reanalysis of the ERA5 meteorological forecast \cite{hersbach2014} with the ground layer being replaced by a measurement from a local anemometer. It has been shown that meteorological forecast wind velocity is consistent with optical turbulence velocity \cite{Osborn2017}. \review{The ERA5 reanalysis has a spatial resolution of \SI{0.25}{\degree} in longitude and latitude and a time resolution of one hour. The error in the wind speed profile can be obtained from the ensemble standard deviation of the forecasts and it was found that the median fractional error for the 35-hour data set was 13\%}. To the best of our knowledge, this is the first time such a hybrid approach has been taken for turbulence monitoring. 

\comment{The coherence time is dominated by strong and fast-moving turbulence and can be estimated from the optical turbulence and velocity profile with

\begin{equation}
\tau _0 = \left(2.914 \left(\frac{2\pi }{\lambda }\right)^2 \int {C_n^2(h) V(h)^{5/3}} \mathrm{d}h \right)^{-3/5},
\end{equation}
where $\lambda$ is the wavelength and $h$ is the altitude of the turbulence.}

\comment{
The coherence angle, or isoplanatic angle, $\theta_0$, can be calculated directly from the vertical turbulence profile, \cn\  from
\begin{equation}
    \theta _0 = \left(2.914 \left(\frac{2\pi }{\lambda }\right)^2  \int C_n^2(h) h^{5/3} \mathrm{d}h \right)^{-3/5},
\end{equation}
demonstrating that stronger high-altitude turbulence leads to a smaller isoplanatic angle.}
\comment{
The Rytov variance, $\sigma_R^2$, defines the magnitude of the received intensity variance after passing through the atmosphere. This is equivalent to the scintillation index for a point in weak turbulence ($\sigma_R^2<0.3$). The Rytov variance can be calculated from the vertical turbulence profile by
\begin{equation}
    \sigma_R^2 = 2.25\left(\frac{2\pi}{\lambda }\right)^{\frac{7}{6}} \int C_n^2(h) h^{5/6} \mathrm{d}h.
\end{equation}}

\review{
\subsection{Scintillation correction}

During the process of computing the theoretical centroid auto-covariance response of the \gls{shimm}, following the analysis in \cite{Butterley2006}, the theoretical centroid covariances between pairs of sub-apertures on the \gls{wfs} are computed numerically. This involves evaluating the spatial covariance of the phase aberration as a result of the turbulence,

\comment{\begin{equation}\label{eq:SLODARCov1}
C^x_{i,j,i',j'} = \iint \langle(\phi(w\mathbf{r}_{i,j})\phi(w\mathbf{r}_{i',j'})\rangle F_x(\mathbf{r}_{i,j}) F_x(\mathbf{r}_{i',j'}) W(\mathbf{r}_{i,j}) W(\mathbf{r}_{i,j}) \mathrm{d}\mathbf{r}_{i,j}  \mathrm{d}\mathbf{r}_{i',j'}
\end{equation}

where $i,j$ and $i',j'$ denote the grid positions of two sub-apertures, $\phi$ the phase aberration relative to the aperture mean, $\mathbf{r}_{i,j}$ is a spatial coordinate defined in units of sub-aperture width with its origin at the centre of sub-aperture $i,j$, $w$ is the width of a sub-aperture, $W(\mathbf{r})$ is the sub-aperture pupil function and $F_x(\mathbf{r})$ is the linear slope function in the $x$ direction, normalised by the sub-aperture pupil function.
The covariance of the phase is evaluated using the following equation,}

\begin{equation}\label{eq:SLODARCov2}
\begin{aligned}
\left\langle\phi(w\mathbf{r}_{i,j})\phi(w\mathbf{r}_{i',j'})\right\rangle = -\frac{1}{2} D_\phi(w\mathbf{x}) + \frac{1}{2}\int W(\mathbf{r}_{i,j}) D_\phi(w\mathbf{x}) \mathrm{d}\mathbf{r}_{i,j} \\
+ \frac{1}{2}\int W(\mathbf{r}_{i',j'}) D_\phi(w\mathbf{x}) \mathrm{d}\mathbf{r}_{i',j'} - \frac{1}{2} \iint W(\mathbf{r}_{i,j}) W(\mathbf{r}_{i',j'}) D_\phi(w\mathbf{x}) \mathrm{d}\mathbf{r}_{i,j} \mathrm{d}\mathbf{r}_{i',j'},
\end{aligned}
\end{equation}

where $(i,j)$ and $(i',j')$ denote the grid positions of two sub-apertures as in \cite{Butterley2006}, $\mathbf{r}_{i,j}$ is a spatial coordinate measured in sub-aperture widths from the centre of sub-aperture $(i,j)$, $w$ is the sub-aperture width, $\phi(w\mathbf{r}_{i,j})$ the phase aberration in subaperture $(i,j)$ relative to the aperture mean, $D_\phi$ is the spatial structure function of phase aberrations, $W(\mathbf{r}_{ij})$ is the sub-aperture pupil function and $\mathbf{x} = (i'-i,j'-j) +\mathbf{r}_{i',j'} - \mathbf{r}_{i,j}$. It is assumed that turbulence measured by the \gls{shimm} is Kolmogorov and therefore the form of the structure function of the phase is

\begin{equation}\label{eq:kolsf}
    D_{\phi}(\mathbf{r}) = 6.88 (r/r_0)^{5/3},
\end{equation}

where $\mathbf{r}$ is a spatial coordinate and $r = |\mathbf{r}|$. However, Eq.~(\ref{eq:kolsf}) does not take into account the effect of scintillation which reduces the contribution of higher-altitude turbulence to the centroid auto-covariance measured by the instrument. As a result, the \gls{slodar} analysis will underestimate the integrated turbulence strength and hence overestimate the value of the coherence length. To correct this we utilise the estimate of the four-layer vertical \cn\  profile and the scintillation-modified Kolmogorov power spectrum of the phase given by \cite{Roddier1981}

\begin{equation}\label{eq:powerspec}
    \phi_{\mathrm{PSD}}(f) = \num{9.7e-3} \left(\frac{2\pi}{\lambda}\right)^2  f^{-11/3} C_n^2(h) \mathrm{d}h \cos^2{\!\left(\pi \lambda h f^2\right)},
\end{equation}

where $f$ is the spatial frequency in the telescope pupil plane and $\lambda$ is the wavelength of the light. Using the following relation, the spatial structure function of the phase may be computed via numerical integration of the phase power spectral density \cite{Jenkins1998}

\begin{equation}\label{eq:sfNumInt}
    D_{\phi}(\mathbf{r}) = 4\pi\! \int\! \mathrm{d}f f \phi_{\mathrm{PSD}}(f) \left[1-J_0\left(2\pi f r\right)\right],
\end{equation}

where $J_0$ is the Bessel function of the first kind of order zero. The integral in Eq.~(\ref{eq:sfNumInt}) can be solved numerically using standard techniques for evaluating improper integrals. Using this scintillation-modified structure function in Eq.~(\ref{eq:SLODARCov2}) and the \cn\ of each layer}, theoretical scintillation-corrected auto-covariance responses are generated  \review{at the effective wavelength of the instrument} for the 4, 12 and 20km layers in the vertical profile. The sum of these responses is then subtracted from the auto-covariance measured by the \gls{shimm} to give the corrected auto-covariance of the \SI{0}{\kilo\meter} ground layer. \review{The SLODAR analysis with the original structure function, Eq.~(\ref{eq:kolsf}), is then applied} to calculate an estimate of the ground layer \cn\  unbiased by scintillation and the integrated turbulence strength of the corrected profile is used to calculate an unbiased estimate of the coherence length. This correction to the \SI{0}{\kilo\meter} ground layer also reduces the bias in the measured coherence time.

\section{Results}
\subsection{Validation in simulation}
\begin{figure}[ht]
    \centering
    \includegraphics[width=0.9\textwidth,trim={0 1cm 0 2.8cm},clip]{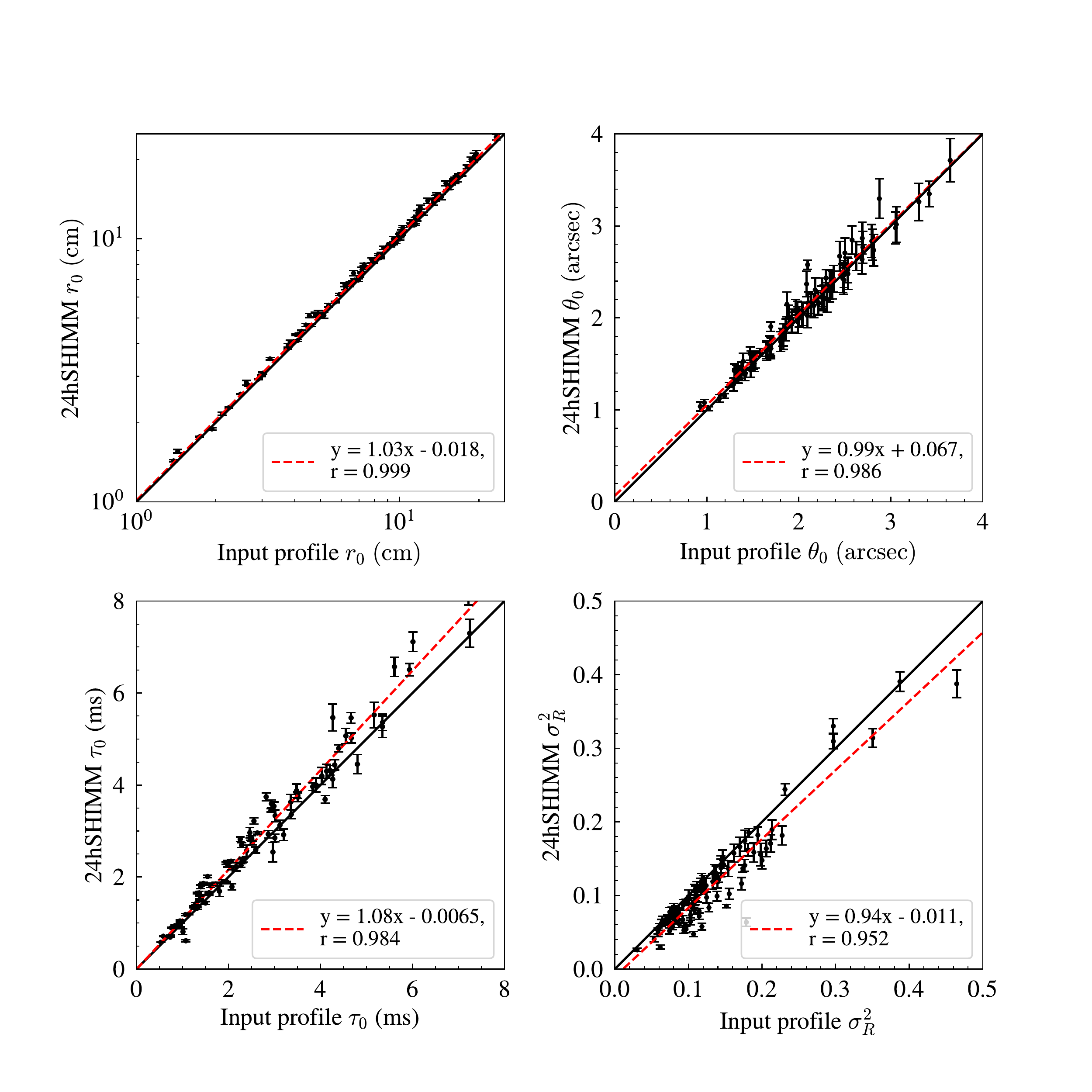}
    \caption{Parameter measurement methods are tested via end to end Monte Carlo simulation on the instrument using measured vertical turbulence profiles from the Stereo-SCIDAR instrument as input. The panels show the coherence length, coherence angle on the top row and the coherence time and Rytov variance on the bottom row. The red dashed line shows the linear best-fit for the data calculated through linear regression, whereas the black solid line indicates the perfect instrument response. r indicates the value of the Pearson correlation coefficient for the data. \review{Simulations were carried out for monochromatic light with a wavelength of \SI{1280}{\nano\meter}, however parameters reported in this figure have been corrected to their values at \SI{500}{\nano\meter}.}}
    \label{fig:simvalid}
\end{figure}

Fig.~\ref{fig:simvalid} shows the turbulence parameters estimated in an end-to-end Monte-Carlo simulation of the methods presented above carried out using the AOtools simulation package \cite{Townson2019}. The input turbulence profiles are taken from the Stereo-SCIDAR database at Paranal, Chile \cite{Osborn2018Profiling}. The simulation is a full physical optics simulation capable of modelling strong turbulence propagation. \review{The simulation assumes monochromatic light at the effective wavelength of the \gls{shimm}, \SI{1280}{\nano\meter}, as this was found to be a very good approximation to running the simulation with multiple wavelengths weighted by the instrument transmission spectrum. The effects of finite exposure time have also been neglected.} To include the stronger low-altitude turbulence expected during the daytime as a result of solar heating, an additional high-strength layer of turbulence has been placed at an altitude of \SI{0}{\meter}. 
The simulations have been carried out for a mean target signal level per spot of 35000$\mathrm{e}^{-}$ and a mean sky background level of 2850 $\mathrm{e}^- \mathrm{pix}^{-1}$. Also included are the shot noise of the source flux and the sky background light based on the mean signal and background flux levels in addition to an RMS detector read noise of 66.5 $\mathrm{e}^- \mathrm{pix}^{-1}$. These values were obtained from \gls{wfs} images taken during the 35-hour experiment in La Palma \review{and give a signal-to-noise ratio of 70.6}. The shot signal is equivalent to that of the faintest target observed during the day and the simulation sky background and the read noise are equivalent to the median values observed during the day. The gain of the camera is $G = 2.01 \mathrm{e}^- \mathrm{ADU}^{-1}$. It should be noted that the RMS read noise here is the quadrature sum of the detector dark noise variance and readout noise variance because a separate measurement of the dark count was impossible as only the dark-subtracted images were saved. \review{For the analysis of simulations, a 2x2 median filter was applied when calculating the centroids to reduce the bias in measurements due to strong noise and the intensities were measured from unfiltered images.}

In Fig.~\ref{fig:simvalid}, the "input profile" atmospheric coherence time is calculated using interpolated wind speed profiles from the stereo-SCIDAR data and a perfect knowledge of this $V(h)$ profile by the \gls{shimm} is assumed. For the reconstructed profile, the wind speed of the \SI{0}{\meter} layer is set as $V(0)$ and for the three atmospheric layers the mean value of $V(h)^{5/3}$ in an \SI{8}{\kilo\meter} bin around each layer is used. This analysis relies on the assumption that these bins are large enough to contain \review{a uniform distribution} of turbulence strength. However the uncertainty in the value of the $V(h)$ depends on the true distribution of $C_n^2(h)$ and $V(h)$ within each bin and so is not included as the \gls{shimm} cannot quantify this, although its effect does contribute to the spread of data points in the figure. 

We see that in general, the turbulence strength is slightly underestimated, particularly in weaker conditions and the strongest conditions. The scintillation correction process described in section 3.4 reduced the bias on measurements of $r_0$ by approximately \review{3\%} and $\tau_0$ by 2\%. The residual bias is primarily due to the pixel sampling in the detector being too large to accurately measure the small centroid motions induced by weak turbulence and speckling of the spots in very strong turbulence reducing the signal-to-noise ratio for centroiding. The coherence angle and Rytov variance however depend solely on the vertical turbulence profile calculated from the intensity covariances. The coherence angle displays little evidence of bias, whereas the Rytov variance appears to be slightly underestimated. From analysing the response of the profile reconstruction to a single turbulent layer, there is evidence that the Rytov variance will be underestimated for profiles with strong low-altitude turbulence and the effect of this is greater on this parameter than the coherence angle due to the relative scaling with height.

\comment{
XXX
isoplanatic angle 2 - C-shape
XXX
}
\comment{
\begin{table*}[htpb]
    \centering
    \begin{tabular}{|c|c|c|}
        \hline
         & gradient & intercept\\
         \hline
         $r_0$  & 1.13  &  -\SI{0.55}{\cm} \\
         $\theta_0$  & 1.07  & -\SI{0.0049}{\arcsecond}  \\
         $\tau_0$  &  1.09 &  -\SI{0.12}{\ms} \\
         $\sigma_R^2$  &  0.85  &  -0.0097 \\
        \hline
    \end{tabular}
    \caption{Gradient and intercept of a line of best fit for a simulation of the atmospheric parameters.}
    \label{tab:linearRegress}
\end{table*}
}

\begin{figure}[hb!]
    \centering
    \includegraphics[width=0.65\textwidth,trim={0 2.5cm 0 4.cm},clip]{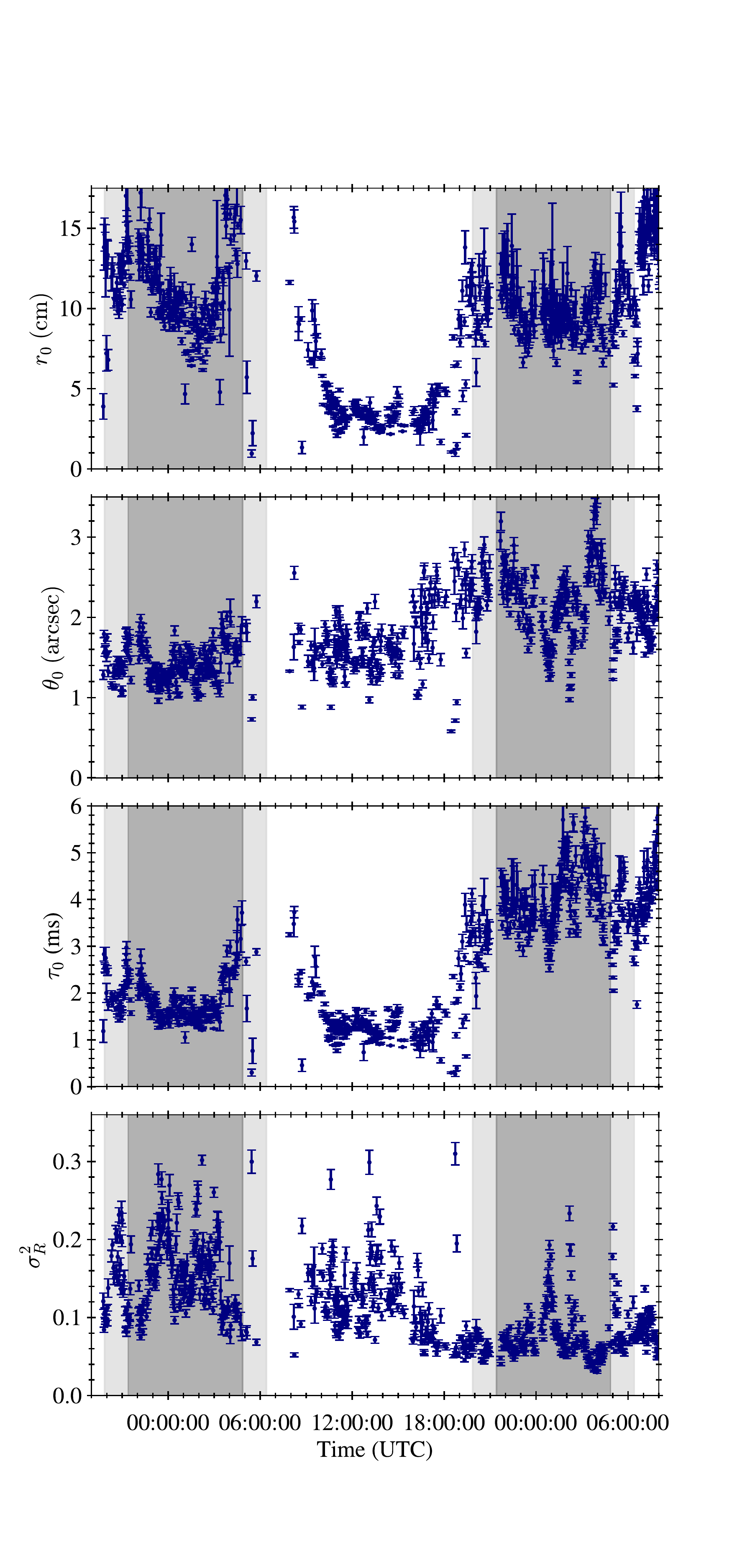}
    \caption{35-hour sequence of turbulence parameters measured at the Observatorio de los Muchachos, La Palma, Spain from the 14th to 16th May 2022. From top to bottom panels show the coherence length, coherence angle, coherence time and Rytov variance. The background colours, dark grey, light grey and white, indicate night, twilight and day respectively. There is a gap around 5am UTC on the first night due to winds in excess of 50 km/hour. Parameters are corrected to zenith and a wavelength of 500~nm.}
    \label{fig:35h}
\end{figure}

\subsection{35-hour continuous monitoring}
Fig.~\ref{fig:35h} shows an example 35-hour sequence of turbulence parameter monitoring at the Roque de Los Muchachos observatory, La Palma, Spain, from the 14th to 16th May 2022. There is a gap from around 5am to 8am UTC on the 15th May due to extremely high levels of wind that temporarily curtailed measurements due to the spot motion as a result of wind shake being larger than the \gls{wfs} field of view. Although the instrument was still running, measurements taken during this time were not reliable and are therefore excluded from the graph. All stars observed during the day were magnitude zero or brighter in the J band. Due to the large number of individual measurements, for presentation purposes only every second data point has been plotted. We see a strong contrast between night and day conditions in $r_0$ and $\tau_0$ but little difference in $\theta_0$ and $\sigma_R^2$. This suggests that solar heating during the daytime increases the surface layer turbulence strength but does not affect higher altitude turbulence.

\section{Conclusion}
\glsresetall
We have demonstrated the first continuous vertical turbulence monitor, capable of working in the day and the night without any modifications or downtime. \review{The instrument uses simultaneous measurements of slopes and intensities in a \gls{wfs} to estimate a four-layer vertical optical turbulence profile, from which the coherence length, angle and Rytov variance are derived, with additional wind speed data from meteorological forecasts used to derive the coherence time. The instrument also implements a method of correcting the ground layer of the measured vertical profile for the effects of scintillation.}

The \gls{shimm} is simple and robust, utilising a well-known and trusted Shack-Hartmann wavefront sensor design. \review{Most of the analysis techniques used in this work are well documented in technical literature and have been validated here using detailed Monte-Carlo  simulations. As such, the on-sky results have not been compared to alternative turbulence monitoring techniques as it was not possible to co-locate instruments in order to match the local turbulence conditions and there were no other instruments available to compare daytime and twilight measurements. This could be investigated in further experiments.}

\comment{\review{Although the analysis techniques used have been validated in simulation, it would be ideal to compare measurements of atmospheric optical turbulence parameters from the \gls{shimm} with simultaneous measurements using an established turbulence monitor such as the stereo-SCIDAR instrument, making sure to co-locate the instruments in order to control for local turbulence effects.}}

The \gls{shimm} is a significant development on previous capabilities. This is an important step forward in the field of ground based optical instrumentation as it will enable site selection, performance optimisation and validation through continuous monitoring for astronomical observatories and free-space optical communications. This development delivers the first instrument that can be used at any time of the day and is small, cheap and portable enough to be used in any location\review{, including urban and sub-urban environments. This instrument will be able to build databases of 24-hour turbulence conditions for current and future optical ground station sites in urban areas, while also providing a wealth of data for assimilation into turbulence forecasting models which are critical for the operation of free space optical communications networks, solar and astronomical observatories. This instrument provides the field with the first truly continuous, day and night turbulence monitor for all but the most demanding of applications.}

\comment{This instrument provides the field with the first truly continuous, day and night turbulence monitor for all but the most demanding of applications.}

\begin{backmatter}
\bmsection{Acknowledgements}
We acknowledge the UK Research and Innovation Future Leaders Fellowship (MR/S035338/1). We thank the  Isaac Newton Group of Telescopes in the Spanish Observatorio del Roque de los Muchachos of the Instituto de Astrofísica de Canarias for support during our test campaigns. RG acknowledges STFC for his studentship funding (project reference 2419794).

\bmsection{Disclosures}
The authors declare no conflicts of interest.

\bmsection{Data availability}
\label{sect:data}
The sequences of turbulence parameters from La Palma are available on reasonable request to the author.
\end{backmatter}


\bibliography{references} 

\end{document}